\documentclass[envcountsame]{llncs}

\usepackage{pslatex}
\usepackage[latin1]{inputenc}
\usepackage{alltt}
\usepackage{graphicx}
\usepackage{subfigure}
\usepackage{dcolumn}
\usepackage{multicol}
\usepackage{amsmath,amssymb}
\usepackage{xspace}
\usepackage{url}
\usepackage{color}

\title{Truly On-The-Fly LTL Model Checking}
\author{Moritz Hammer\inst{1} \and Alexander Knapp\inst{1}
  \and Stephan Merz\inst{2}}
\institute{%
  Institut f\"ur Informatik,
  Ludwig-Maximilians-Universit{\"a}t M{\"u}nchen\\
  \email{\{Moritz.Hammer,Alexander.Knapp\}@pst.ifi.lmu.de}
\and%
  INRIA Lorraine, LORIA, Nancy\\
  \email{Stephan.Merz@loria.fr}
}

\DeclareMathOperator{\dom}{dom}

\newcommand{\nat}{\mathbb{N}}

\newcommand{\move}{\rightarrow}

\newcommand{\LWAASpin}{\textsc{LwaaSpin}\xspace}
\newcommand{\LTLBA}{\textsc{ltl2ba}\xspace}
\newcommand{\SPIN}{\textsc{Spin}\xspace}
\newcommand{\oom}{\multicolumn{1}{r|}{o.o.m.}}
\newcommand{\oot}{\multicolumn{1}{r|}{o.o.t.}}

\renewcommand{\phi}{\varphi}
\renewcommand{\AA}{\mathcal{A}}
\newcommand{\BB}{\mathcal{B}}
\newcommand{\LL}{\mathcal{L}}
\newcommand{\TT}{\mathcal{T}}
\newcommand{\VV}{\mathcal{V}}

\renewcommand{\implies}{\Rightarrow}

\newcommand{\suffix}[2]{#1|_{#2}}
\newcommand{\true}{\textbf{\upshape true}}

\newcommand{\nex}{\mathop{\textbf{X}}}
\newcommand{\alw}{\mathop{\textbf{G}}}
\newcommand{\eve}{\mathop{\textbf{F}}}
\newcommand{\until}{\mathrel{\mathbf{U}}}
\newcommand{\release}{\mathrel{\mathbf{V}}}

\newcolumntype{.}{D{.}{.}{-1}}

\newcommand{\code}[1]{\texttt{\def\{{\char123}\def\}{\char125}\def\^{\char94}\def\_{\char95}#1}}
\newenvironment{CODE}{%
  \begin{alltt}\small\def\{{\char123}\def\}{\char125}\def\^{\char94}\def\_{\char95}}{%
  \end{alltt}}
\newcommand{\kw}[1]{\textbf{#1}}

\begin{document}
\maketitle

\begin{abstract}
  We propose a novel algorithm for automata-based LTL model checking
  that interleaves the construction of the generalized Büchi automaton
  for the negation of the formula and the emptiness check. Our
  algorithm first converts the LTL formula into a linear weak
  alternating automaton; configurations of the alternating automaton
  correspond to the locations of a generalized Büchi automaton, and a
  variant of Tarjan's algorithm is used to decide the existence of an
  accepting run of the product of the transition system and the
  automaton. Because we avoid an explicit construction of the Büchi
  automaton, our approach can yield significant improvements in
  runtime and memory, for large LTL formulas. The algorithm has been
  implemented within the \SPIN{} model checker, and we present
  experimental results for some benchmark examples.
\end{abstract}

\section{Introduction}
\label{sec:introduction}

The automata-based approach to linear-time temporal logic (LTL) model
checking reduces the problem of deciding whether a formula $\varphi$
holds of a transition system $\TT$ into two subproblems: first, one
constructs an automaton $\AA_{\lnot\varphi}$ that accepts precisely
the models of $\lnot\varphi$. Second, one uses graph-theoretical
algorithms to decide whether the product of $\TT$ and
$\AA_{\lnot\varphi}$ admits an accepting run; this is the case if and
only if $\varphi$ does not hold of $\TT$. On-the-fly
algorithms~\cite{courcoubetis:memory-efficient} avoid an explicit
construction of the product and are commonly used to decide the second
problem. However, the construction of a non-deterministic Büchi (or
generalized Büchi) automaton $\AA_{\lnot\varphi}$ is already of
complexity exponential in the length of $\varphi$, and several
algorithms have been
suggested~\cite{daniele:improved,etessami:buechi,fritz:buchi,gastin:fast,somenzi:efficient,tauriainen:translating}
that improve on the classical method for computing Büchi
automata~\cite{gerth:simple}.  Still, there are applications, for
example when verifying liveness properties over predicate
abstractions~\cite{kesten:liveness}, where the construction of
$\AA_{\lnot\varphi}$ takes a significant fraction of the overall
verification time. The relative cost of computing $\AA_{\lnot\varphi}$
is particularly high when $\varphi$ does not hold of $\TT$, because
acceptance cycles are often found rather quickly when they exist.

In this paper we suggest an algorithm for LTL model checking that
interleaves the construction of (a structure equivalent to) the
automaton and the test for non-emptiness.  Technically, the input to
our algorithm is a transition system $\TT$ and a linear weak
alternating automaton (LWAA, alternatively known as a very weak
alternating automaton) corresponding to $\lnot\varphi$.  The size of
the LWAA is linear in the length of the LTL formula, and the time for
its generation is insignificant. It can be considered as a symbolic
representation of the corresponding generalized Büchi automaton (GBA).
LWAA have also been employed as an intermediate format in the
algorithms suggested by Gastin and Oddoux~\cite{gastin:fast},
Fritz~\cite{fritz:buchi}, and Schneider~\cite{schneider:yet-another}.
Our main contribution is the identification of a class of ``simple''
LWAA whose acceptance criterion is defined in terms of the sets of
locations activated during a run, rather than the standard criterion
in terms of automaton transitions. To explore the product of the
transition system and the configuration graph of the LWAA, we employ a
variant of Tarjan's algorithm to search for a strongly connected
component that satisfies the automaton's acceptance condition.

We have implemented the proposed algorithm as an alternative
verification method in the \SPIN{} model
checker~\cite{holzmann:spin-book}, and we discuss some implementation
options and report on experimental results. Our implementation is
available for download at
\url{http://www.pst.ifi.lmu.de/projekte/lwaaspin/}.

\section{LTL and linear weak alternating automata}
\label{sec:ltl-lwaa}

We define alternating $\omega$-automata, especially LWAA, and present
the translation from propositional linear-time temporal logic LTL to
LWAA. Throughout, we assume a fixed finite set $\VV$ of atomic
propositions.

\subsection{Linear weak alternating automata}
\label{sec:lwaa}

We consider automata that operate on temporal structures, i.e.\ 
$\omega$-sequences of valuations of $\VV$.  Alternating automata
combine the existential branching mode of non-deterministic automata
(i.e., choice) with its dual, universal branching, where several
successor locations are activated simultaneously. We present the
transitions of alternating automata by associating with every location
$q \in Q$ a propositional formula $\delta(q)$ over $\VV$ and $Q$. For
example, we interpret
\[
  \delta(q_1)\ \ =\ \ 
  (v \land q_2 \land (q_1 \lor q_3)) \lor
  (\lnot w \land q_1) \lor
  w
\]
as asserting that if location $q_1$ is currently active and the
current input satisfies $v$ then the automaton should simultaneously
activate the locations $q_2$ and either $q_1$ or $q_3$. If the input
satisfies $\lnot w$ then $q_1$ should be activated. If the input
satisfies $w$ then no successor locations need to be activated from
$q_1$.  Otherwise (i.e., if the input satisfies $\lnot v$), the
automaton blocks because the transition formula can not be satisfied.
At any point during a run, a set of automaton locations (a
\emph{configuration}) will be active, and transitions are required to
satisfy the transition formulas of all active locations.  Locations $q
\in Q$ may only occur positively in transition formulas: locations
cannot be inhibited.  We use the following generic definition of
alternating $\omega$-automata:

\begin{definition}\label{def:alternating}
  An \emph{alternating $\omega$-automaton} is a tuple $\AA = (Q, q_0,
  \delta, Acc)$ where
  \begin{itemize}
  \item $Q$ is a finite set (of locations) where $Q \cap \VV =
    \emptyset$,
  \item $q_0 \in Q$ is the initial location,
  \item $\delta: Q \rightarrow \BB(Q \cup \VV)$ is the transition
    function that associates a propositional formula $\delta(q)$ with
    every location $q \in Q$; locations in $Q$ can only occur
    positively in $\delta(q)$,
  \item and $Acc \subseteq Q^{\omega}$ is the acceptance condition.
  \end{itemize}
\end{definition}

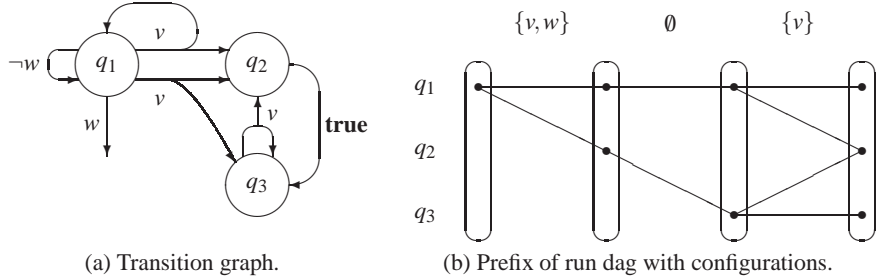
\begin{figure}[tp]
  \hspace*{\fill}
  \subfigure[Transition graph.]{%
    \label{fig:auto-trans}
    \makebox[40mm]{%
      \unitlength 1mm
      \begin{picture}(28,28)
        \put(4,20){\circle{8}}\put(4,20){\makebox(0,0){$q_1$}}
        \put(24,20){\circle{8}}\put(24,20){\makebox(0,0){$q_2$}}
        \put(24,4){\circle{8}}\put(24,4){\makebox(0,0){$q_3$}}
        \put(7.7,22){\vector(1,0){12.6}}\put(11,24){\makebox(0,0){$v$}}
        \put(10,24.2){\oval(12,4.4)[rb]}
        \put(10,24.2){\oval(12,8)[t]}\put(4,24.2){\vector(0,-1){0}}
        \put(7.7,18){\vector(1,0){12.6}}\put(11,16){\makebox(0,0)[t]{$v$}}
        \qbezier(12,18)(15,18)(21,7)\put(21,7){\vector(2,-3){0}}
        \put(4,15.8){\vector(0,-1){8}}\put(3,12){\makebox(0,0)[r]{$w$}}
        \put(0.2,20){\oval(8,4)[l]}
        \put(0.2,18){\vector(1,0){0}}
        \put(-5,20){\makebox(0,0)[r]{$\neg w$}}
        \put(24,7.8){\oval(4,8)[t]}
        \put(24,11.8){\vector(0,1){4}}
        \put(26,7.8){\vector(0,-1){0}}
        \put(26,14){\makebox(0,0)[t]{$v$}}
        \put(28.2,12){\oval(8,16)[r]}
        \put(28.2,4){\vector(-1,0){0}}
        \put(33,12){\makebox(0,0)[l]{$\true$}}
      \end{picture}
  }}
  \hfill\hfill
  \subfigure[Prefix of run dag with configurations.]{%
    \label{fig:run-dag}
    \unitlength 1.7mm
    \begin{picture}(35,15)
      \put(0,10){\makebox(0,0)[l]{$q_1$}}
      \put(0,5){\makebox(0,0)[l]{$q_2$}}
      \put(0,0){\makebox(0,0)[l]{$q_3$}}
      \put(10,15){\makebox(0,0){$\{v,w\}$}}
      \put(20,15){\makebox(0,0){$\emptyset$}}
      \put(30,15){\makebox(0,0){$\{v\}$}}
      \put(5,10){\circle*{.6}}
      \put(5,10){\line(1,0){10}}
      \put(5,10){\line(2,-1){10}}
      \put(15,10){\circle*{.6}}
      \put(15,5){\circle*{.6}}
      \put(15,10){\line(1,0){10}}  
      \put(15,5){\line(2,-1){10}}
      \put(25,10){\circle*{.6}}
      \put(25,0){\circle*{.6}}
      \put(25,10){\line(1,0){10}}
      \put(25,10){\line(2,-1){10}}
      \put(25,0){\line(2,1){10}}
      \put(25,0){\line(1,0){10}}
      \put(35,10){\circle*{.6}}
      \put(35,5){\circle*{.6}}
      \put(35,0){\circle*{.6}}
      \put(5,5){\oval(2,14)}
      \put(15,5){\oval(2,14)}
      \put(25,5){\oval(2,14)}
      \put(35,5){\oval(2,14)}
    \end{picture}
  }
  \hspace*{\fill}
  \caption{Visualization of alternating automata and run dags.}
  \label{fig:visual}
\end{figure}

When the transition formulas $\delta(q)$ are written in disjunctive
normal form, the alternating automaton can be visualized as a
hypergraph. For example, Fig.~\ref{fig:auto-trans} shows an
alternating $\omega$-automaton and illustrates the above transition
formula. We write $q \move q'$ if $q$ may activate $q'$, i.e.\ if $q'$
appears in $\delta(q)$.

Runs of an alternating $\omega$-automaton over a temporal structure
$\sigma = s_0 s_1 \ldots$ are not just sequences of locations but give
rise to trees, due to universal branching. However, different copies
of the same target location can be identified, and we obtain a more
economical dag representation as illustrated in
Fig.~\ref{fig:run-dag}: the vertical ``slices'' of the dag represent
configurations that are active before reading the next input state.

We identify a set and the Boolean valuation that makes true precisely
the elements of the set. For example, we say that the sets
$\{v,w,q_2,q_3\}$ and $\{w\}$ satisfy the formula $\delta(q_1)$ above.
For a relation $r \subseteq S \times T$, we denote its domain by
$\dom(r)$.  We denote the image of a set $A \subseteq S$ under $r$ by
$r(A)$; for $x \in S$ we sometimes write $r(x)$ for $r(\{x\})$.

\begin{definition}\label{def:run-dag}
  Let $\AA = (Q, q_0, \delta, Acc)$ be an alternating
  $\omega$-automaton and $\sigma = s_0 s_1 \ldots$, where $s_i
  \subseteq \VV$, be a temporal structure. A \emph{run dag} of $\AA$
  over $\sigma$ is represented by the $\omega$-sequence $\Delta = e_0
  e_1 \ldots$ of its edges $e_i \subseteq Q \times Q$. The
  configurations $c_0 c_1 \ldots$ of $\Delta$, where $c_i \subseteq
  Q$, are inductively defined by $c_0 = \{q_0\}$ and $c_{i+1} =
  e_i(c_i)$.  We require that for all $i \in \nat$, $\dom(e_i)
  \subseteq c_i$ and that for all $q \in c_i$, the valuation $s_i \cup
  e_i(q)$ satisfies $\delta(q)$. A \emph{finite run dag} is a finite
  prefix of a run dag.
  
  A \emph{path} in a run dag $\Delta$ is a (finite or infinite)
  sequence $\pi = p_0 p_1 \ldots$ of locations $p_i \in Q$ such that
  $p_0 = q_0$ and $(p_i, p_{i+1}) \in e_i$ for all $i$.
  A run dag $\Delta$ is \emph{accepting} iff $\pi \in Acc$ holds for
  all infinite paths $\pi$ in $\Delta$. The \emph{language} $\LL(\AA)$
  is the set of words that admit some accepting run dag.
\end{definition}

Because locations do not occur negatively in transition formulas
$\delta(q)$, it is easy to see that whenever $s_i \cup X$ satisfies
$\delta(q)$ for some set $X$ of locations, then so does $s_i \cup Y$
for any superset $Y$ of $X$. However, the dag resulting from replacing
$X$ by $Y$ will have more paths, making the acceptance condition
harder to satisfy. It is therefore enough to consider only run dags
that arise from minimal models of the transition formulas w.r.t.\ the
states of the temporal structure, activating as few successor
locations as possible.

LWAA are alternating $\omega$-automata whose accessibility relation
determines a partial order: $q'$ is reachable from $q$ only if $q'$ is
smaller or at most equal to $q$. We are interested in LWAA with a
co-Büchi acceptance condition:

\begin{definition}\label{def:vwaa}
  A (co-Büchi) linear weak alternating automaton $\AA = (Q, q_0,
  \delta, F)$ is a tuple where $Q$, $q_0$, and $\delta$ are as in
  Def.~\ref{def:alternating} and $F \subseteq Q$ is a set of locations,
  such that
  \begin{itemize}
  \item the relation $\preceq_{\AA}$ defined by $q' \preceq_{\AA} q$
    iff $q \move^* q'$ is a partial order on $Q$ and
  \item the acceptance condition is given by
    \[
      Acc = \{p_0 p_1 \ldots \in Q^{\omega} : 
              p_i \in F \text{ for only finitely many } i \in \nat\}.
    \]
  \end{itemize}
\end{definition}

In particular, the hypergraph of the transitions of an LWAA does not
contain cycles other than self-loops, and run dags of LWAA do not
contain ``rising edges'' as in Fig.~\ref{fig:visual}. It follows that
every infinite path eventually remains stable at some location $q$,
and the acceptance condition requires that $q \notin F$ holds for that
``limit location''. LWAA characterize precisely the class of star-free
$\omega$-regular languages, which correspond to first-order definable
$\omega$-languages and therefore also to the languages definable by
propositional LTL formulas~\cite{rohde:alternating,thomas:languages}.

\subsection{From LTL to LWAA}
\label{sec:ltl-vwaa}

Formulas of LTL (over atomic propositions in $\VV$) are built using
the connectives of propositional logic and the temporal operators
$\nex$ (next) and $\until$ (until). They are interpreted over a
temporal structure $\sigma = s_0s_1\ldots \in (2^{\VV})^{\omega}$ as
follows; we write $\suffix{\sigma}{i}$ to denote the suffix $s_i
s_{i+1} \ldots$ of $\sigma$ from state $s_i$:
\[
\renewcommand{\arraystretch}{1.2}
\begin{array}{l@{\text{\ \ \ iff\ \ \ }}l@{\qquad\qquad}l@{\text{\ \ \ iff\ \ \ }}l}
  \sigma \models p
  & p \in s_0 &
  \sigma \models \varphi \land \psi
  & \sigma \models \varphi \text{\ \ and\ \ } \sigma \models \psi\\
  \sigma \models \lnot\varphi &
  \sigma \not\models \varphi &
  \sigma \models \nex\varphi 
  & \suffix{\sigma}{1} \models \varphi\\
  \sigma \models \varphi \until \psi
  & \multicolumn{3}{l}{
    \text{for some $i \in \nat$, $\suffix{\sigma}{i} \models \psi$
      and for all $j<i$, $\suffix{\sigma}{j} \models \varphi$}}
\end{array}
\]

We freely use the standard derived operators of propositional logic
and the following derived temporal connectives:
\[
\begin{array}{r@{\ \ }c@{\ \ }l@{\qquad}l}
  \eve\varphi & \equiv & \true \until \varphi
  & \text{(eventually $\varphi$)}\\
  \alw\varphi & \equiv & \lnot\eve\lnot\varphi
  & \text{(always $\varphi$)}\\
  \varphi \release \psi & \equiv & \lnot(\lnot\varphi \until \lnot\psi)
  & \text{($\varphi$ releases $\psi$)}
\end{array}
\]

\begin{figure}[tp]
  \subfigure[Transition formulas of $\AA_{\varphi}$]{%
    \label{fig:vwaa-trans}%
    \renewcommand{\arraystretch}{1.2}%
    \begin{tabular}{@{}|*{2}{@{\ \ }c@{\ \ }|}}
      \hline
      location $q$ & $\delta(q)$\\
      \hline\hline
      $q_{\psi}$ ($\psi$ a literal) & $\psi$\\
      \hline
      $q_{\psi \land \chi}$ & 
      $\delta(q_{\psi}) \land \delta(q_{\chi})$\\
      \hline
      $q_{\psi \lor \chi}$ & 
      $\delta(q_{\psi}) \lor \delta(q_{\chi})$\\
      \hline
      $q_{\nex\psi}$ & $q_{\psi}$\\
      \hline
      $q_{\psi \until \chi}$ &
      $\delta(q_{\chi}) \lor (\delta(q_{\psi}) \land q_{\psi \until \chi})$\\
      \hline
      $q_{\psi \release \chi}$ &
      $\delta(q_{\chi}) \land (\delta(q_{\psi}) \lor q_{\psi \release \chi})$\\
      \hline
    \end{tabular}}
  \hfill
  \subfigure[$\AA_{\alw\eve p}$]{%
    \label{fig:vwaa-GFp}%
    \unitlength 1mm
    \begin{minipage}{20\unitlength}
    \begin{picture}(20,35)
      \put(5,30){\circle{10}}
      \put(5,30){\makebox(0,0){$\alw\eve p$}}
      \put(5,10){\circle{10}}
      \put(5,10){\circle{9}}
      \put(5,10){\makebox(0,0){$\eve p$}}
      \put(9,30){\oval(9,6)[r]}\put(9,33){\vector(-1,0){0}}
      \put(14.5,30){\makebox(0,0)[l]{$p$}}
      \put(9,10){\oval(9,6)[r]}\put(9,13){\vector(-1,0){0}}
      \put(14.5,10){\makebox(0,0)[l]{$\lnot p$}}
      \put(5,5){\vector(0,-1){5}}
      \put(6,2.5){\makebox(0,0)[l]{$p$}}
      \put(5,25.5){\oval(4,9)[b]}\put(7,25.5){\vector(0,1){0}}
      \put(5,21){\vector(0,-1){6}}
      \put(6,19){\makebox(0,0)[l]{$\lnot p$}}
    \end{picture}
    \end{minipage}}
  \hfill
  \subfigure[$\AA_{p \until (q \until r)}$]{%
    \label{fig:vwaa-pUqUr}
    \unitlength 1mm
    \begin{minipage}{33\unitlength}
    \begin{picture}(30,35)(6,0)
      \put(11,10){\circle{10}}
      \put(11,10){\circle{9}}
      \put(11,10){\makebox(0,0){$q \until r$}}
      \put(16,30){\oval(20,10)}
      \put(16,30){\oval(18,8)}
      \put(16,30){\makebox(0,0){$p \until (q \until r)$}}
      \put(25,30){\oval(10,6)[r]}\put(25,33){\vector(-1,0){0}}
      \put(31,30){\makebox(0,0)[l]{$p \land \lnot r$}}
      \put(21,25){\vector(0,-1){6}}
      \put(22,22){\makebox(0,0)[l]{$r$}}
      \put(11,25){\vector(0,-1){10}}
      \put(10,22){\makebox(0,0)[r]{$q \land \lnot r$}}
      \put(15,10){\oval(10,6)[r]}\put(15,13){\vector(-1,0){0}}
      \put(21,10){\makebox(0,0)[l]{$q \land \lnot r$}}
      \put(11,5){\vector(0,-1){5}}
      \put(12.6,2.5){\makebox(0,0){$r$}}
    \end{picture}
    \end{minipage}}
  \caption{Translation of LTL formulas into LWAA.}
  \label{fig:ltl-vwaa}
\end{figure}
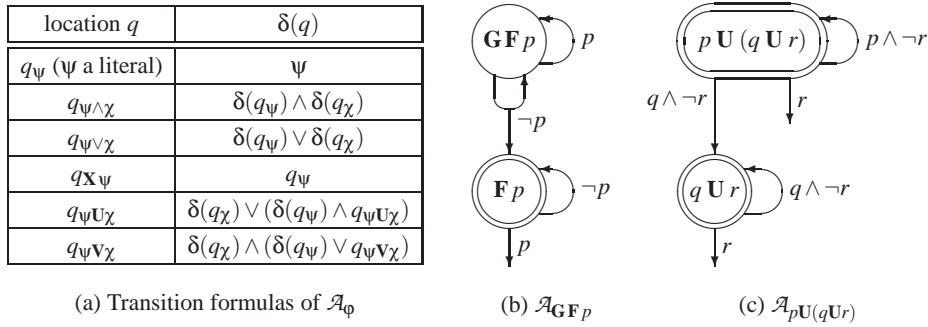

An LTL formula $\varphi$ can be understood as defining the language
\[
  \LL(\varphi)\ =\ 
  \{\sigma \in (2^{\VV})^{\omega} : \sigma \models \varphi\},
\]
and the automata-theoretic approach to model checking builds on this
identification of formulas and languages, via an effective
construction of automata $\AA_{\varphi}$ accepting the language
$\LL(\varphi)$. The definition of an LWAA $\AA_{\varphi}$ is
particularly simple~\cite{muller:weak}: without loss of generality, we
assume that LTL formulas are given in negation normal form (i.e.,
negation is applied only to propositions), and therefore include
clauses for the dual operators $\lor$ and $\release$.  The automaton
is $\AA_{\varphi} = (Q, q_{\varphi}, \delta, F)$ where $Q$ contains a
location $q_{\psi}$ for every subformula $\psi$ of $\varphi$, with
$q_{\varphi}$ being the initial location.  The transition formulas
$\delta(q_{\psi})$ are defined in Fig.~\ref{fig:vwaa-trans}; in
particular, LTL operators are simply decomposed according to their
fixpoint characterizations. The set $F$ of co-final locations consists
of all locations $q_{\psi \until \chi} \in Q$ that correspond to
``until'' subformulas of $\varphi$. It is easy to verify that the
resulting automaton $\AA_{\varphi}$ is an LWAA: for any locations
$q_{\psi}$ and $q_{\chi}$, the definition of $\delta(q_{\psi})$
ensures that $q_{\psi} \move q_{\chi}$ holds only if $\chi$ is a
subformula of $\psi$. Correctness proofs for the construction can be
found in~\cite{muller:weak,vardi:alternating}; conversely,
Rohde~\cite{rohde:alternating} and Löding and
Thomas~\cite{loeding:alternating} prove that for every LWAA $\AA$
there is an LTL formula $\varphi_{\AA}$ such that $\LL(\varphi_{\AA})
= \LL(\AA)$.

The number of subformulas of an LTL formula $\varphi$ is linear in the
length of $\varphi$, and therefore so is the size of $\AA_{\varphi}$.
However, in practice the automaton should be minimized further.
Clearly, unreachable locations can be eliminated.  Moreover, whenever
there is a choice between activating sets $X$ or $Y$ of locations
where $X \subseteq Y$ from some location $q$, the smaller set $X$
should be preferred, and $Y$ should be activated only if $X$ cannot
be.  As a simple example, we can define \( \delta(q_{\eve p}) = p \lor
(\lnot p \land q_{\eve p}) \text{ instead of } \delta(q_{\eve p}) = p
\lor q_{\eve p} \).

Figure~\ref{fig:ltl-vwaa} shows two linear weak alternating automata
obtained from LTL formulas by applying this construction (the
locations in $F$ are indicated by double circles).

Further minimizations are less straightforward. Because the automaton
structure closely resembles the structure of the LTL formula,
heuristics to minimize the LTL
formula~\cite{etessami:buechi,somenzi:efficient} are important. Fritz
and Wilke~\cite{fritz:simulation} discuss more elaborate optimizations
based on simulation relations on the set $Q$ of locations.

\section{Deciding language emptiness for LWAA}
\label{sec:emptiness-vwaa}

In general, it is nontrivial to decide language emptiness for
alternating $\omega$-automata, due to their intricate combinatorial
structure: a configuration consists of a set of automaton locations
that have to ``synchronize'' on the current input state during a
transition to a successor configuration. The standard approach is
therefore based on a translation to non-deterministic Büchi automata,
for which emptiness can be decided in linear time.  Unfortunately,
this translation is of exponential complexity.

Linear weak alternating automata have a simpler combinatorial
structure: the transition graph contains only trivial cycles, and
therefore a run dag is non-accepting only if it contains a path that
ends in a self-loop at some location $q \in F$. This observation gives
rise to the following non-emptiness criterion for LWAA, which is
closely related to Theorem~2 of~\cite{gastin:fast}:

\begin{theorem}\label{thm:vwaa-empty}
  Assume that $\AA = (Q,q_0,\delta,F)$ is an LWAA. Then $\LL(\AA)
  \neq \emptyset$ if and only if there exists a finite run dag $\Delta
  = e_0 e_1 \ldots e_n$ with configurations $c_0 c_1 \ldots c_{n+1}$
  over a finite sequence $s_0 \ldots s_n$ of states and some $k \leq
  n$ such that
  \begin{enumerate}
  \item $c_k = c_{n+1}$ and
  \item for every $q \in F$, one has $(q,q) \notin e_j$ for some $j$
    where $k \leq j \leq n$.
  \end{enumerate}
\end{theorem}
\begin{proof}
  ``If'': Consider the infinite dag $\Delta' = e_0 \ldots e_{k-1} (e_k
  \ldots e_n)^{\omega}$. Because $c_k = c_{n+1}$, it is obvious that
  $\Delta'$ is a run dag over $\sigma = s_0 \ldots s_{k-1} (s_k \ldots
  s_n)^{\omega}$; we now show that $\Delta'$ is accepting. Assume, to
  the contrary, that $\pi = p_0 p_1 \ldots$ is some infinite path in
  $\Delta'$ such that $p_i \in F$ holds for infinitely many $i \in
  \nat$. Because $\AA$ is an LWAA, there exists some $m \in \nat$ and
  some $q \in Q$ such that $p_i = q$ for all $i \geq m$. It follows
  that $(q,q) \in e_i$ holds for all $i \geq m$, which is impossible
  by assumption (2) and the construction of $\Delta'$.  Therefore,
  $\Delta'$ must be accepting, and $\LL(\AA) \neq \emptyset$.

  ``Only if'': Assume that $\sigma = s_0 s_1 \ldots \in \LL(\AA)$, and
  let $\Delta' = e_0 e_1 \ldots$ be some accepting run dag of $\AA$
  over $\sigma$. Since $Q$ is finite, $\Delta'$ can contain only
  finitely many different configurations $c_0, c_1, \ldots$, and there
  is some configuration $c \subseteq Q$ such that $c_i = c$ for
  infinitely many $i \in \nat$. Denote by $i_0 < i_1 < \ldots$ the
  $\omega$-sequence of indexes such that $c_{i_j} = c$. If there were
  some $q \in F$ such that $q \in e_j(q)$ for all $j \geq i_0$
  (implying in particular that $q \in c_j$ for all $j \geq i_0$ by
  Def.~\ref{def:run-dag}) then $\Delta'$ would contain an infinite
  path ending in a self-loop at $q$, contradicting the assumption that
  $\Delta'$ is accepting. Therefore, for every $q \in F$ there must be
  some $j_q \geq i_0$ such that $(q,q) \notin e_{j_q}$. Choosing
  $k=i_0$ and $n = i_m-1$ for some $m$ such that $i_m > j_q$ for all
  (finitely many) $q \in F$, we obtain a finite run dag $\Delta$ as
  required.
  \qed
\end{proof}

Observe that Thm.~\ref{thm:vwaa-empty} requires to inspect the
\emph{transitions} of the dag and not just the configurations. In
fact, a run dag may well be accepting although some location $q \in F$
is contained in all (or almost all) configurations. For example,
consider the LWAA for the formula $\alw\nex\eve p$: the location
$q_{\eve p}$ will be active in every run dag from the second
configuration onward, even if the run dag is accepting. We now
introduce a class of LWAA for which it is enough to inspect the
configurations.

\begin{definition}\label{def:simple}
  An LWAA $\AA = (Q,q_0,\delta,F)$ is \emph{simple} if for all $q \in
  F$, all $q' \in Q$, all states $s \subseteq \VV$, and all $X,Y
  \subseteq Q$ not containing $q$, if $s \cup X \cup \{q\} \models
  \delta(q')$ and $s \cup Y \models \delta(q)$ then $s \cup X \cup Y
  \models \delta(q')$.
\end{definition}

In other words, if a co-final location $q$ can be activated from some
location $q'$ for some state $s$ while it can be exited during the
same transition, then $q'$ has an alternative transition that avoids
activating $q$, and this alternative transitions activates only
locations that would anyway have been activated by the joint
transitions from $q$ and $q'$. For simple LWAA, non-emptiness can be
decided on the basis of the visited configurations alone, without
memorizing the graph structure of the run dag.

\begin{theorem}\label{thm:simple-empty}
  Assume that $\AA = (Q,q_0,\delta,F)$ is a simple LWAA. Then
  $\LL(\AA) \neq \emptyset$ if and only if there exists a finite run
  dag $\Delta = e_0 e_1 \ldots e_n$ with configurations $c_0 c_1
  \ldots c_{n+1}$ over a finite sequence $s_0 \ldots s_n$ of states
  and some $k \leq n$ such that
  \begin{enumerate}
  \item $c_k = c_{n+1}$ and
  \item for every $q \in F$, one has $q \notin c_j$ for some $j$ where
    $k \leq j \leq n$.
  \end{enumerate}
\end{theorem}
\begin{proof}
  ``If'': The assumption $q \notin c_j$ and the requirement that
  $\dom(e_j) \subseteq c_j$ imply that $(q,q) \notin e_j$, and
  therefore $\LL(\AA) \neq \emptyset$ follows using
  Thm.~\ref{thm:vwaa-empty}.
  
  ``Only if'': Assume that $\LL(\AA) \neq \emptyset$, obtain a finite
  run dag $\Delta$ satisfying the conditions of
  Thm.~\ref{thm:vwaa-empty}, and let $l = n-k+1$ denote the length of
  the loop. ``Unwinding'' $\Delta$, we obtain an infinite run dag $e_0
  e_1 \ldots$ over the temporal structure $s_0 s_1 \ldots$ whose edges
  are $e_i = e_{k+((i-k) \bmod l)}$ for $i>n$, and similarly for the
  states $s_i$ and the configurations $c_i$.  W.l.o.g.\ we assume that
  the dag contains no unnecessary edges, i.e.\ that for all $e_i \in
  \Delta$, $(q,q') \in e_i$ holds only if $q \move q'$.
  
  We inductively construct an infinite run dag $\Delta' = e_0' e_1'
  \ldots$ with configurations $c_0' c_1' \ldots$ such that $c_i'
  \subseteq c_i$ as follows: let $c_0' = c_0$ and for $i<k$, let $e_i'
  = e_i$ and $c_{i+1}' = c_{i+1}$. For $i \geq k$, assume that $c_i'$
  has already been defined.  Let $F_i$ denote the set of $q \in c_i'
  \cap F$ such that $(q,q) \notin e_i$ but $q \in e_i(c_i')$, and for
  any $q \in F_i$ let $Q'_q$ denote the set of locations $q' \in c_i'$
  such that $(q',q) \in e_i$ and let $Y_q = e_i(q)$. Because $\AA$ is
  simple, it follows that $s_i \cup (e_i(q') \setminus \{q\}) \cup Y_q
  \models \delta(q')$, for all $q \in F_i$ and $q' \in Q'_q$. We let
  $e_i'$ be obtained from the restriction of $e_i$ to $c_i'$ by
  deleting all edges $(q',q)$ for $q \in F_i$ and adding edges
  $(q',q'')$ for all $q' \in Q'_q$ and $q'' \in Y_q$, for $q \in F_i$.
  Clearly, this ensures that $c_{i+1}' \subseteq c_{i+1}$ holds for
  the resulting configuration and that $c_{i+1}' \cap F_i =
  \emptyset$.
  
  For any $q \in F_i$, the definition of an LWAA and the assumption
  that $q \notin Y_q$ ensure that $q'' \prec_{\AA} q$ holds for all
  $q'' \in Y_q$, as well as $q \preceq_{\AA} q'$ for all $q' \in
  Q'_q$.  In particular, we must have $q'' \neq q'$ for all $q'' \in
  Y_q$ and $q' \in Q'_q$, and therefore $e_i'$ does not contain more
  self loops than $e_i$: for all $p \in Q$, we have $(p,p) \in e_i'$
  only if $(p,p) \in e_i$.
  
  Consequently, $\Delta'$ is an accepting infinite run dag such that
  for every $q \in F$ there exists some $j \geq k$ such that $q \notin
  c_j'$. It now suffices to pick some $n \geq k$ satisfying the
  conditions of the theorem; such an $n$ exists because $F$ is finite
  and $\Delta'$ can contain only finitely many different
  configurations.
  \qed
\end{proof}

\begin{figure}[tp]
\begin{tabular}{@{}cc@{}}
    \unitlength 1mm
    \begin{minipage}{20\unitlength}
    \begin{picture}(20,35)
      \put(5,30){\circle{10}}
      \put(5,30){\makebox(0,0){$\alw\eve p$}}
      \put(5,10){\circle{10}}
      \put(5,10){\circle{9}}
      \put(5,10){\makebox(0,0){$\eve p$}}
      \put(9,30){\oval(9,6)[r]}\put(9,33){\vector(-1,0){0}}
      \put(14.5,30){\makebox(0,0)[l]{$p$}}
      \put(9,10){\oval(9,6)[r]}\put(9,13){\vector(-1,0){0}}
      \put(14.5,10){\makebox(0,0)[l]{$\true$}} 
      \put(5,5){\vector(0,-1){5}}
      \put(6,2.5){\makebox(0,0)[l]{$p$}}
      \put(5,25.5){\oval(4,9)[b]}\put(7,25.5){\vector(0,1){0}}
      \put(5,21){\vector(0,-1){6}}
      \put(6,19){\makebox(0,0)[l]{$\true$}} 
    \end{picture}
    \end{minipage}
& \begin{minipage}{10cm}
  \setlength{\unitlength}{1.05pt}
  \begin{tabular}[t]{@{}l@{}}
\begin{picture}(270,35)
\multiput(40,25)(30,0){7}{\circle*{3}}
\multiput(70,10)(30,0){6}{\circle*{3}}
\put(20,25){\makebox(0,0){$\alw \eve p$}}
\put(20,10){\makebox(0,0){$\eve p$}}
\put(55,32){\makebox(0,0){$\{p\}$}}
\put(85,32){\makebox(0,0){$\{p\}$}}
\put(115,32){\makebox(0,0){$\emptyset$}}
\put(145,32){\makebox(0,0){$\{p\}$}}
\put(175,32){\makebox(0,0){$\{p\}$}}
\put(205,32){\makebox(0,0){$\emptyset$}}
\put(235,32){\makebox(0,0){$\{p\}$}}
\put(255,25){\makebox(0,0){$\ldots$}}
\put(255,10){\makebox(0,0){$\ldots$}}
\multiput(40,25)(30,0){7}{\put(2,0){\vector(1,0){26}}}
\multiput(40,25)(30,0){7}{\put(2,-1){\vector(2,-1){26}}}
\put(100,10){\put(2,0){\vector(1,0){26}}}
\put(160,10){\put(2,0){\vector(1,0){26}}}
\put(190,10){\put(2,0){\vector(1,0){26}}}
\put(70,10){\put(2,-1){\vector(2,-1){13}}}
\put(130,10){\put(2,-1){\vector(2,-1){13}}}
\put(220,10){\put(2,-1){\vector(2,-1){13}}}
\end{picture}

  \\[4ex]
\begin{picture}(270,35)
\multiput(40,25)(30,0){7}{\circle*{3}}
\put(70,10){\circle*{3}}
\put(130,10){\circle*{3}}
\put(190,10){\circle*{3}}
\put(220,10){\circle*{3}}
\put(20,25){\makebox(0,0){$\alw \eve p$}}
\put(20,10){\makebox(0,0){$\eve p$}}
\put(55,32){\makebox(0,0){$\{p\}$}}
\put(85,32){\makebox(0,0){$\{p\}$}}
\put(115,32){\makebox(0,0){$\emptyset$}}
\put(145,32){\makebox(0,0){$\{p\}$}}
\put(175,32){\makebox(0,0){$\{p\}$}}
\put(205,32){\makebox(0,0){$\emptyset$}}
\put(235,32){\makebox(0,0){$\{p\}$}}
\put(255,25){\makebox(0,0){$\ldots$}}
\put(255,10){\makebox(0,0){$\ldots$}}
\multiput(40,25)(30,0){7}{\put(2,0){\vector(1,0){26}}}
\put(40,25){\put(2,-1){\vector(2,-1){26}}}
\put(100,25){\put(2,-1){\vector(2,-1){26}}}
\put(160,25){\put(2,-1){\vector(2,-1){26}}}
\put(190,25){\put(2,-1){\vector(2,-1){26}}}
\put(190,10){\put(2,0){\vector(1,0){26}}}
\put(70,10){\put(2,-1){\vector(2,-1){13}}}
\put(130,10){\put(2,-1){\vector(2,-1){13}}}
\put(220,10){\put(2,-1){\vector(2,-1){13}}}
\end{picture}

  \end{tabular}
\end{minipage}
\end{tabular}
  \caption{Illustration of the construction of Thm.~\protect{~\ref{thm:simple-empty}}.}
  \label{fig:simple-rundags}
\end{figure}
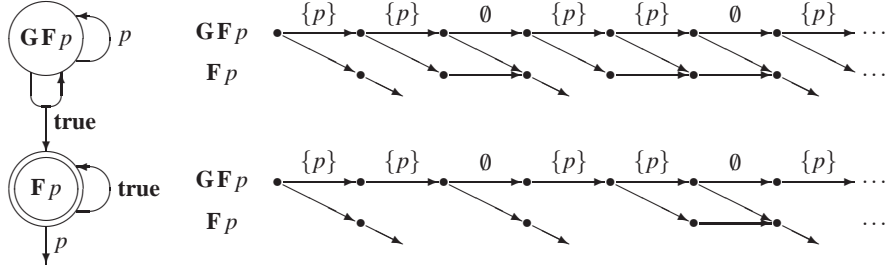

Fig.~\ref{fig:simple-rundags} illustrates two accepting run dags for a
simple LWAA: the dag shown above satisfies the criterion of
Thm.~\ref{thm:vwaa-empty} although the co-final location corresponding
to $\eve p$ remains active from the second configuration onward. The
dag shown below is the result of the transformation described in the
proof, and indeed the location $\eve p$ is infinitely often inactive.

We now show that the LWAA $\AA_{\varphi}$ for an LTL formula $\varphi$
is simple provided $\varphi$ does not contain subformulas $\nex(\chi
\until \chi')$.  Such subformulas are easily avoided because $\nex$
distributes over $\until$.  
Actually, our implementation exploits the commutativity of $\nex$ with
all LTL connectives to rewrite formulas such that no other temporal
operators are in the scope of $\nex$; this is useful for preliminary
simplifications at the formula level.  Also, the transformations
described at the end of Sect.~\ref{sec:ltl-vwaa} ensure that the LWAA
remains simple.

%

\begin{theorem}\label{thm:ltl-simple}
  For any LTL formula $\varphi$ that does not contain any subformula
  $\nex(\chi \until \chi')$, the automaton $\AA_{\varphi}$ is a simple
  LWAA.
\end{theorem}
\begin{proof}
  Let $\AA_{\varphi} = (Q,q_{\varphi},\delta,F)$ and assume that $q
  \in F$, $q' \in Q$, and $X,Y \subseteq Q$ are as in
  Def.~\ref{def:simple}, in particular $s \cup X \cup \{q\} \models
  \delta(q')$ and $s \cup Y \models \delta(q)$.  The proof is by
  induction on $\psi$ where $q' = q_{\psi}$.

  \begin{description}
  \item[$\psi \equiv (\lnot)v:$] $\delta(q') = \psi$, so we must have
    $s \models \delta(q')$, and the assertion $s \cup X \cup Y \models
    \delta(q')$ follows trivially.
  \item[$\psi \equiv \chi \otimes \chi',\ \otimes \in \{\land,\lor\}:$]
    $\delta(q') = \delta(q_{\chi}) \otimes \delta(q_{\chi'})$, and the
    assertion follows easily from the induction hypothesis.
  \item[$\psi \equiv \nex\chi:$] $\delta(q') = q_{\chi}$, and by
    assumption $\chi$ is not an $\until$ formula, so $q_{\chi} \notin
    F$. In particular, $q_{\chi} \neq q$, and so the assumption $s
    \cup X \cup \{q\} \models \delta(q')$ implies that $s \cup X
    \models \delta(q')$, and the assertion $s \cup X \cup Y \models
    \delta(q')$ follows by monotonicity.
  \item[$\psi \equiv \chi \until \chi':$] $\delta(q') =
    \delta(q_{\chi'}) \lor (\delta(q_{\chi}) \land q')$. In case $s
    \cup X \cup \{q\} \models \delta(q_{\chi'})$, the induction
    hypothesis implies $s \cup X \cup Y \models \delta(q_{\chi'})$,
    hence also $s \cup X \cup Y \models \delta(q')$.
  
    If $s \cup X \cup \{q\} \models \delta(q_{\chi}) \land q'$, we
    consider two cases: if $q=q'$ then $s \cup Y \models \delta(q')$
    holds by assumption. Moreover, $s \cup X \cup Y \models
    \delta(q_{\chi})$ holds by induction hypothesis, and the assertion
    follows.
    
    Otherwise, we must have $q' \in X$. Again, $s \cup X \cup Y
    \models \delta(q_{\chi})$ follows from the induction hypothesis,
    and since $q' \in X$ it follows that $s \cup X \cup Y \models
    \delta(q_{\chi}) \land q'$.
  \item[$\psi \equiv \chi \release \chi':$] $\delta(q') =
    \delta(q_{\chi'}) \land (\delta(q_{\chi}) \lor q')$. In
    particular, $s \cup X \cup \{q\} \models \delta(q_{\chi'})$, and
    we obtain $s \cup X \cup Y \models \delta(q_{\chi'})$ by induction
    hypothesis.

    If $s \cup X \cup \{q\} \models \delta(q_{\chi})$, we similarly
    obtain $s \cup X \cup Y \models \delta(q_{\chi})$. Otherwise, note
    that $q \neq q'$ because $q \in F$ and $q' \notin F$ (since it
    is not an $\until$ formula). Therefore, we must have $s \cup X
    \models q'$, and a fortiori $s \cup X \cup Y \models q'$,
    completing the proof.
    \qed
  \end{description}
\end{proof}

Let us note in passing that simple LWAA are as expressive as LWAA,
i.e.\ they also characterize the class of star-free $\omega$-regular
languages: from~\cite{loeding:alternating,rohde:alternating} we know
that for every LWAA $\AA$ there is an LTL formula $\varphi_{\AA}$ such
that $\LL(\varphi_{\AA}) = \LL(\AA)$. Since $\nex$ distributes over
$\until$, $\varphi_{\AA}$ can be transformed into an equivalent
formula $\varphi'$ of the form required in Thm.~\ref{thm:ltl-simple},
and $\AA_{\varphi'}$ is a simple LWAA accepting the same language as
$\AA$.

\section{Model checking algorithm}
\label{sec:mc-algorithm}

We describe a model checking algorithm based on the nonemptiness
criterion of Thm.~\ref{thm:simple-empty}, and we discuss some design
decisions encountered in our implementation. The algorithm has been
integrated within the LTL model checker \SPIN{}, and we present some
results that have been obtained on benchmark examples.

\subsection{Adapting Tarjan's algorithm}
\label{sec:mc-overview}

Theorem~\ref{thm:simple-empty} contains the core of our model checking
algorithm: given the simple LWAA $\AA_{\lnot\varphi}$ corresponding to
the negation $\lnot\varphi$ of the property to be verified, we explore
the product of the transition system $\TT$ and the graph of
configurations of $\AA_{\lnot\varphi}$, searching for a strongly
connected component that satisfies the acceptance condition.  In fact,
in the light of Thm.~\ref{thm:simple-empty} a simple LWAA $\AA$ can
alternatively be viewed as a symbolic representation of a GBA whose
locations are sets of locations of $\AA$, and that has an acceptance
condition per co-final location of $\AA$.

\begin{figure}[tp]
  \begin{CODE}\fontsize{8}{.8\baselineskip}
  \selectfont 
  \kw{procedure} Visit(s, C):
    \kw{let} c = (s,C) \kw{in}
      inComp[c] := \kw{false}; root[c] := c; labels[c] := \(\emptyset\);
      cnt[c] := cnt; cnt := cnt+1; seen := seen \(\cup\) \{c\};
      push(c, stack);
      \kw{forall} c' = (s',C') in Succ(c) \kw{do}
        \kw{if} c' \(\notin\) seen \kw{then} Visit(s',C') \kw{end if};
        \kw{if} \(\lnot\)inComp[c'] \kw{then}
          \kw{if} cnt[root[c']] < cnt[root[c]] \kw{then}
            labels[root[c']] := labels[root[c']] \(\cup\) labels[root[c]];
            root[c] := root[c']
          \kw{end if};
          labels[root[c]] := labels[root[c]] 
                             \(\cup\) (f_lwaa \(\setminus\) C); // f_lwaa \(\equiv\) co-final locations
          \kw{if} labels[root[c]] = f_lwaa \kw{then} \kw{raise} Good\_Cycle \kw{end if};
        \kw{end if};
      \kw{end forall};
      \kw{if} root[c]=c \kw{then}
        \kw{repeat}
          d := pop(stack);
          inComp[d] := \kw{true};
        \kw{until} d=c; 
      \kw{end if};
    \kw{end let};
  \kw{end} Visit;

  \kw{procedure} Check:
    stack := empty; seen := \(\emptyset\); cnt := 0;
    Visit(init_ts, \{init_lwaa\});    // start with initial location
  \kw{end} Check;  \end{CODE}
  \caption{LWAA-based model checking algorithm.}
  \label{fig:tarjan}
\end{figure}

The traditional CVWY algorithm~\cite{courcoubetis:memory-efficient}
for LTL model checking based on Büchi automata has been generalized
for GBA by Tauriainen~\cite{tauriainen:nested}, but we find it easier
to adapt Tarjan's algorithm~\cite{tarjan:depth} for finding strongly
connected components in graphs.  Figure~\ref{fig:tarjan} gives a
pseudo-code representation of our algorithm. The depth-first search
operates on pairs $(s,C)$ where $s$ is a state of the transition
system and $C$ is a configuration of the LWAA. Given a pair $c =
(s,C)$, the call to \code{Succ} computes the set
$\mathit{succ}_{\TT}(s) \times \mathit{succ}_{\AA}(s,C)$ containing
all pairs $c' = (s',C')$ of successor states $s'$ of the transition
system and successor configurations $C'$ of the LWAA, i.e.\ those $C'$
which satisfy $s \cup C' \models \delta(q)$ for all $q \in C$.
Tarjan's algorithm assigns a so-called root candidate \code{root} to
each node of the graph, which is the oldest node on the stack known to
belong to the same SCC.

In model checking, we are not so much interested in actually computing
SCCs: it is sufficient to verify that the acceptance criterion of
Thm.~\ref{thm:simple-empty} is met for some strongly connected
subgraph (SCS). To do so, we associate a \code{labels} field with the
root candidate of each SCC that accumulates the locations $q \in F$
that have been found absent in some pair $(s,C)$ contained in the SCC.
Whenever \code{labels} is found to contain all co-final states of the
LWAA (denoted by \code{f\_lwaa}), the SCS must be accepting and the
search is aborted. Note that we need to maintain two stacks: one for
the depth-first search recursion, and one for identifying SCCs.

If an accepting SCS is found, we also want to produce a
counter-example, and Tarjan's algorithm is less convenient for this
purpose than the CVWY algorithm whose recursion stack contains the
counter-example once a cycle has been detected. In our case, neither
the recursion stack nor the SCC stack represent a complete
counter-example. A counter-example can still be obtained by traversing
the nodes of an accepting SCS that have already been visited, without
re-considering the transition system. We add two pointers to our node
representation in the SCC stack, representing ``backward'' and
``forward'' links that point to the pair from which the current node
was reached and to the oldest pair on the stack that is a successor of
the current pair.  Indeed, one can show that the subgraph of nodes on
the SCC stack with neighborhood relation
\[
  \{(c,c') : c'=\mathit{forward}(c)\ \text{or}\ c = \mathit{backward}(c') \}
\]
also forms an SCS of the product graph. A counter-example can now be
produced by enforcing a visit to all the pairs that satisfy some
acceptance condition.

\subsection{Computation of successor configurations}
\label{sec:successors}

The efficient generation of successor configurations in
$\mathit{succ}_{\AA}(s,C)$ is a crucial part of our algorithm. Given a
configuration $C \subseteq Q$ of the LWAA and a state $s$ of the
transition system (which we identify with a valuation of the
propositional variables), we need to compute the set of all $C'$ such
that $s \cup C' \models \delta(q)$ holds for all $q \in C$. Moreover,
we are mainly interested in finding minimal successor configurations.

An elegant approach towards computing successor configurations makes
use of BDDs~\cite{bryant:bdds}. In fact, the transitions of an LWAA
can be represented by a single BDD. The set of minimal successor
configurations is obtained by conjoining this BDD with the BDD
representations of the state $s$ and the source configuration $C$, and
then extracting the set of all satisfying valuations of the resulting
BDD. Some experimentation convinced us, however, that the resulting
BDDs become too big for large LTL formulas. Alternatively, one can
store BDDs representing $\delta(q)$ for each location $q$ and form the
conjunction of all $\delta(q)$ for $q \in C$. Again, this approach
turned out to consume too much memory.

We finally resorted to using BDDs only as a representation of
configurations.  To do so, we examine the hyperedges of the transition
graph of the LWAA, which correspond to the clauses of the disjunctive
normal form of $\delta(q)$. For every location $q \in C$, we compute
the disjunction of its enabled transitions, and then take the
conjunction over all locations in $C$. We thus obtain
\[
  \mathit{succ}_{\AA}(s,C)\ =\ 
  \bigwedge_{q \in C} \big(
    \bigvee_{t \in \mathit{enabled}(s,q)} (t \setminus \VV)
  \big)
\]
as the BDD representing the set of successor configurations, where
$\mathit{enabled}(s,q)$ denotes the set of enabled transitions of $q$
for state $s$, i.e.\ those transitions $t$ for which $s \cup Q \models
t$. Although this requires pre-computing a potentially exponentially
large set of transitions, this approach appears to be fastest for
BDD-based calculation of successor nodes.

We compare this approach to a direct calculation of successor
configurations that stores them as a sorted list, which is pruned to
remove non-minimal successors. Although the pruning step is of
quadratic complexity in our implementation (it could be improved to
$O(n \log n)$ time), experiments showed that it pays off handsomely
because fewer nodes need to be explored in the graph search.

\subsection{Adapting Spin}
\label{sec:adapting-spin}

Either approach to computing successors works best if we can
efficiently determine the set of enabled transitions of an LWAA
location. One way to do this is to generate C source code for a given
LWAA and then use the CPU arithmetics. The \SPIN{} model checker
employs a similar approach, albeit for Büchi automata, and this is one
of reasons why we adapted it to use our algorithm.

\SPIN{}~\cite{holzmann:spin,holzmann:spin-book}, is generally
considered as one of the fastest and most complete tools for protocol
verification. For a given model (written in Promela) and Büchi
automaton (called ``never-claim''), it generates C sources that are
then compiled to produce a model-specific model checker. \SPIN{} also
includes a translation from LTL formulas to Büchi automata, but for
our comparisons we used the \LTLBA{} tool due to Gastin and
Oddoux~\cite{gastin:fast}, which is faster by orders of magnitude for
large LTL formulas.

Our adaptation, called \LWAASpin{}, adds the generation of LWAA to
\SPIN{}, and modifies the code generation to use Tarjan's algorithm
and on-the-fly calculation of successor configurations. This involved
about 150 code changes, and added about 2600 lines of code. \SPIN{}
includes elaborate optimizations, such as partial-order reduction,
that are independent of the use of non-deterministic or alternating
automata and that can therefore be used with our implementation as
well.  We have not yet adapted \SPIN{}'s optimizations of memory usage
such as bitstate hashing to our algorithm, although we see no obstacle
in principle to do so.

\subsection{Experimental results}
\label{sec:results}

Geldenhuys and Valmari~\cite{geldenhuys:tarjan} have recently proposed
to use Tarjan's algorithm, but for non-deterministic Büchi automata,
and we have implemented their algorithm for comparison.  We have not
been able to reproduce their results indicating that Tarjan's
algorithm outperforms the CVWY algorithm on nondeterministic Büchi
automata (their paper does not indicate which implementation of CVWY
was used).  In our experiments, both algorithms perform head-to-head
on most examples. We now describe the results for the implementation
based on LWAA.

For most examples, the search for an accepting SCS in the product
graph is slower than the runtime of the model checker produced by
\SPIN{} after \LTLBA{} has generated the Büchi automaton. However, our
algorithm can be considerably faster than generating the Büchi
automaton and then checking the emptiness of the product automaton,
for large LTL formulas.  However, note that both \SPIN and our
implementation use unguided search, and we can thus not exactly
compare single instances of satisfiable problems.

Large LTL formulas are not as common as one might expect. \SPIN's
implementation of the CVWY algorithm can handle weak fairness of
processes directly; such conditions do not have to be added to the LTL
formula to be verified. We present two simple and scalable examples:
the dining philosophers problem and a binary semaphore protocol.

For the dining philosophers example, we want to verify that if every
philosopher holds exactly one fork infinitely often, then philosopher
$1$ will eventually eat:
\[
  \alw\eve\mathit{hasFork}_1 \land \ldots \land \alw\eve\mathit{hasFork}_n
  \ \implies\ \alw\eve \mathit{eat}_1
\]

The model \texttt{dinphil}$n$ denotes the situation where all $n$
philosophers start with their right-hand fork, which may lead to a
deadlock. The model \texttt{dinphil}$n$\texttt{i} avoids the deadlock
by letting the $n$-th philosopher start with his left-hand fork.

For the binary semaphore example we claim that if strong fairness is
ensured for each process, all processes will eventually have been in
their critical section:
\[ 
   (\alw\eve\mathit{canenter}_1 \implies \alw\eve\mathit{enter}_1)
   \land \ldots \land
   (\alw\eve\mathit{canenter}_n \implies \alw\eve\mathit{enter}_n)
   \ \implies\ \eve \mathit{allcrit}
\]

By \texttt{sfgood}$n$, we denote a constellation with $n$ processes
and strong fairness assumed for each of them, while \texttt{sfbad}$n$
denotes the same constellation, except with weak fairness for process
$p_n$, which will allow the process to starve.


\begin{table}[tp]
\begin{center}
\begin{tabular}{@{}|l|c||.|.|.|.||.|.|.|@{}}\hline
\multicolumn{1}{|c|}{Problem} & \multicolumn{1}{c||}{Counter-} &
\multicolumn{4}{c||}{\SPIN} & \multicolumn{3}{c|}{\LWAASpin} \\ \cline{3-9}
&  \multicolumn{1}{c||}{example} & \multicolumn{1}{c|}{\texttt{ltl2ba}} & \multicolumn{1}{c|}{\texttt{spin}} &
\multicolumn{1}{c|}{\texttt{gcc}} & \multicolumn{1}{c||}{\texttt{pan}} &
\multicolumn{1}{c|}{\texttt{lwaaspin}} & \multicolumn{1}{c|}{\texttt{gcc}} &
\multicolumn{1}{c|}{\texttt{pan}} \\ \hline \hline
dinphil6 & yes & 0.431 & 0.019 & 0.601 & 0.079 & 0.019 & 0.579 & 0.163\\ \hline
dinphil8 & yes & 35.946 & 0.02 & 0.671 & 0.133 & 0.027 & 0.818 & 0.166 \\ \hline
dinphil10& yes & 3611.724 & 0.025 & 0.767 & 1.642 & 0.057 & 1.899 & 0.170\\ \hline
dinphil12& yes& \oot & & & & 0.141 & 6.644 & 0.206 \\ \hline
dinphil14& yes& & & & & 0.499 & 28.082 & 0.431 \\ \hline
dinphil15& yes& & & & & 0.972 & \oom & \\ \hline
dinphil6i& no & 0.431 & 0.024 & 0.639 & 0.244 &  0.020 & 0.616 & 0.569 \\ \hline
dinphil8i& no & 35.946 & 0.021 & 0.711 & 7.309 & 0.028 & 0.861 & 20.177 \\ \hline
dinphil10i&no& 3611.724 & 0.025 & 0.807 & 722.874 & 0.070 & 2.623 & 623.760 \\ \hline
dinphil11i&no& \oot &        &       &       & 0.099 & 3.438 & \oom \\ \hline \hline
sfbad6 & yes & 1.904 & 0.912 & 7.284 & 0.025 & 0.066 & 2.211 & 1.312 \\ \hline
sfbad7 & yes & 27.674 & 42.525 & \oom &  & 0.179 & 7.423 & 7.848 \\ \hline
sfbad8 & yes &        &        &      &  & 0.784 & 43.472 & 7.000 \\ \hline
sfbad9 & yes &        &        &      &  & 2.627 & \oom & \\ \hline
sfgood6 & no & 2.292 & 17.329 & 27.608 & 2.193 & 0.064 & 2.227 & 2.540 \\
\hline
sfgood7 & no & 36.306 & 417.485 & \oom &  & 0.357 & 8.214 & 15.940 \\ \hline
sfgood8 & no &        &         &      &  & 0.718 & 42.688 & 140.130\\ \hline
sfgood9 & no &        &         &      &  & 2.634 & \oom & \\ \hline
\end{tabular}
\end{center}
\caption{Comparison of \SPIN and \LWAASpin (BDD-less successor calculation)
  \label{tab:results}}
\end{table}

Table~\ref{tab:results} contains timings (in seconds) for the
different steps of the verification process for \SPIN
4.1.1 and for our \LWAASpin implementation. \SPIN
requires successive invocations of \texttt{ltl2ba}, \texttt{spin},
\texttt{gcc} and \texttt{pan}; \textsc{lwaaspin} combines the first
two stages. The times were measured on an Intel
Pentium\textsuperscript{\textregistered}~4, 3.0~GHz computer with 1GB
main memory running Linux and without other significant process
activity. Entries ``o.o.t.'' indicate that the computation did not
finish within 2~hours, while ``o.o.m.''  means ``out of memory''.

We can see that most of the time required by \SPIN is spent on
preparing the \texttt{pan} model checker, either by calculating the
non-deterministic Büchi automata for the dining philosophers, or by
handling the large automata sources for the binary semaphore example.
\LWAASpin{} significantly reduces the time taken for pre-processing.

\begin{table}[tp]
\begin{center}
\begin{tabular}{@{}|l||.|.||r|r|r|r|r|@{}}\hline 
\multicolumn{1}{|c||}{Problem} & \multicolumn{2}{c||}{Successor calculation} &
\multicolumn{2}{c|}{LWAA} & \multicolumn{2}{c|}{B\"uchi} & \multicolumn{1}{c|}{States} \\ \cline{2-7}
& \multicolumn{1}{c|}{BDD} & \multicolumn{1}{c||}{direct} &
\multicolumn{1}{c|}{Locations} & \multicolumn{1}{c|}{Transitions} &
\multicolumn{1}{c|}{Locations} & \multicolumn{1}{c|}{Transitions} &
\multicolumn{1}{c|}{seen} \\ \hline \hline
dinphil6 & 0.834 & 0.761  & 10 & 207 & 8 & 36 & 105 \\ \hline
dinphil8 & 1.194 & 1.011  & 12 & 787 & 10& 55 & 119 \\ \hline
dinphil10& 2.803 & 2.126  & 14 & 3095& 12& 78 & 133 \\ \hline
dinphil6i & 1.291& 1.205  & 10 & 207 & 8 & 36 & 46165 \\ \hline
dinphil8i &21.802&21.021  & 12 & 787 & 10& 55 & 1.2 $\cdot$ 10$^\text{6}$ \\ \hline
dinphil10i&643.006& 626.453 &  14 & 3095& 12& 78 & 1.5 $\cdot$ 10$^\text{7}$ \\
\hline \hline
sfbad6  & 16.664 & 3.589 & 26  & 4140& 252 & 1757 & 137882 \\ \hline
sfbad7  & 354.874 & 15.461 & 30&16435& 1292& 8252 & 597686 \\ \hline
sfgood6 & 32.261 & 4.831 &  26 & 4139& 972 & 5872 & 221497 \\ \hline
sfgood7 & 115.539 & 24.511 & 30&16434& 3025& 23391&872589 \\ \hline
\end{tabular}
\end{center}
\caption{Comparison of successor calculation, and
  sizes of the automata.\label{tab:automata}}
\vspace*{-4ex}
\end{table}

The sizes of the generated automata are indicated in Tab.~\ref{tab:automata}.
``States seen'' denotes the number of distinct states (of the product
automaton) encountered by \LWAASpin using the direct successor configuration
calculation approach.  It should be noted that the Büchi automata for the
dining philosophers example are very small compared to the size of the
formula, and are in fact linear; even for the \texttt{dinphil10i} case, the
automaton contains only 12~locations.  This is not true for the semaphore
example: the Büchi automaton for \texttt{sfgood7} contains 3025 locations and
23391 transitions.  Still, one advantage of using \LTLBA{} is that a Büchi
automaton that has been computed once can be stored and reused; this could
reduce the overall verification time for the dining philosophers example where
the same formula is used for both the valid and the invalid model.

We can draw two conclusions from our data: first, the preprocessing by
\texttt{lwaaspin} uses very little time because we do not have to
calculate the Büchi automaton (although strictly speaking our
implementation is also exponential because it transforms the
transition formulas into disjunctive normal form). This makes up for
the usually inferior performance of our \texttt{pan} version.  It also
means that we can at least start a model checking run, even for very
large LTL formulas, in the hope of finding a counter-example. Second,
we can check larger LTL formulas.  Ultimately, we encounter the same
difficulties as \SPIN during both the \texttt{gcc} and the
\texttt{pan} phases; after all, we are confronted with a
PSPACE-complete problem. The pre-processing phase could be further
reduced by avoiding the generation of an exponential number of
transitions in the C sources, postponing more work to the \texttt{pan}
executable.  Besides, the bitstate hashing technique as implemented in
\SPIN{}~\cite{holzmann:analysis} could also be applied to Tarjan's
algorithm.

Table~\ref{tab:automata} also compares the two approaches to computing
successor configurations described in Sect.~\ref{sec:successors}. The
BDD-based approach appears to be less predictable and never
outperforms the direct computation, but further experience is
necessary to better understand the tradeoff.

\section{Conclusion and further work}
\label{sec:conclusion}

We have presented a novel algorithm for the classical problem of LTL
model checking. It uses an LWAA encoding of the LTL property as a
symbolic representation of the corresponding GBA, which is effectively
generated on the fly during the state space search, and never has to
be stored explicitly. By adapting the \SPIN model checker to our
approach, we validate that, for large LTL formulas, the time gained by
avoiding the expensive construction of a non-deterministic Büchi
automaton more than makes up for the runtime penalty due to the
implicit GBA generation during model checking, and this advantage does
not appear to be offset by the simplifications applied to the
intermediate automata by algorithms such as \LTLBA. However, we do not
yet really understand the relationship between minimizations at the
automaton level and the local optimizations applied in our search.

We believe that our approach opens the way to verifying large LTL
formulas by model checking. Further work should investigate the
possibilities that arise from this opportunity, such as improving
techniques for software model checking based on predicate abstraction.
Also, our implementation still leaves room for performance
improvements. In particular, the LWAA should be further minimized, the
representation of transitions could be reconsidered, and the memory
requirements could be reduced by clever coding techniques.

\bibliographystyle{plain}
\bibliography{bib}

\end{document}